\documentclass[a4paper,11pt]{article}
\usepackage{jheppub} 
\usepackage{amsmath}
\usepackage{array}
\usepackage{amssymb}
\usepackage{enumitem} 
\usepackage{lineno}
\usepackage{booktabs}
\usepackage{multicol}
\usepackage{multirow}
\usepackage{braket}
\usepackage{epstopdf}
\usepackage{natbib}

\title{\boldmath A Satellite \texorpdfstring{$N=2$}{N=2} Superparticle in Extra Dimensions.}

\author{F. E. A. de Souza, M. O. Tahim, R. I. O. Júnior, I. M. Macêdo.}
\affiliation{Universidade Estadual do Ceará, Faculdade de Educação, Ciências e Letras do Sertão Central\\
Quixadá-CE, Brazil.}

\emailAdd{francisco.emmanoel@uece.br}

\abstract{In this work we discuss the issue of localization of $N=2$ spinning particles. More specifically, we show that we can not confine the spinning particle within the Randall-Sundrum scenario. We argue that this result directly affects studies related to localization of $p$-form fields. We show that, due to the non confinement of the superparticle, we can not localize $p$-forms on the membrane.}

\begin{document}
\maketitle
\flushbottom
\section{Introduction}

There are two paths to  physically start describing antisymmetric tensor fields. The first one is by the direct work with the field representations according to the space-time dimension and group theory arguments. In this case we know that $p$-forms exists and we construct, for instance, duality relations between them \cite{Nakahara:2003nw,Zee:2016fuk}. The second one is by the quantization of a $N=2$ supersymmetric particle, where a spinning particle plays the main role. The quantization procedure imposes some constraints in this model, including the interpretation through $p$-form fields and yet realization/non realization of gravitational backgrounds \cite{Howe:1989vn}. In physics of extra dimensions, it is natural to ask about the field content of a given model. In the case of braneworlds, $p$-forms have a lot of importance: we can explain space-time torsion in $D=4$ \cite{Mukhopadhyaya:2002jn} or discuss axion physics \cite{Svrcek:2006yi}. Despite these, other fields can give hints about the existence of extra-dimensions \cite{Mukhopadhyaya:2007jn,Germani:2004jf}.

Studies of $p$-form field localization in the Randall-Sundrum scenario are clear: $p$-forms can be localized. This can be achieved through several mechanisms. In order to find this result it is necessary to consider Randall-Sundrum backgrounds with dilatonic couplings, where it is argued that, for instance in the specific case of $1$-form gauge fields, the dilaton coupling contribute with warp factors circumventing problems with the natural conformal symmetry of this model \cite{Tahim:2008ka,Alencar:2010mi}. By another hand, geometric couplings, where geometric tensors are taken into account, can give rise to localization \cite{Alencar:2018cbk}. In another way, it was found that the Kaluza Klein (KK) modes of the $p$-form fields satisfy a Schröedinger-like equation: the characters of the KK modes are decided by the behavior of the effective potential, which depends on the form parameters and the background geometry. The localization will depends specifically of the $p$-brane model \cite{Fu:2012sa, Fu:2016vaj}. A recent work regarded the coupling with gravity directly with kinetic $p$-form terms, obtaining again results with localization \cite{Lu:2024gmx}.

However, there are works where the issue of localization of pure particle models gives the opposite answer: there is no natural particle confinement to a braneworld, unless the dilaton coupling is properly adressed. Despite this case, in particular, for the case of the $N=1$ spinning particle model in $D=5$, the particle is confined in another region, different from the brane's position, such a characteristic which renders the non-confinement of the superparticle \cite{Souza:2019jqz}. In order to add the spin degrees of freedom at the classical level, the idea is to introduce Grassmann algebra in quantum mechanics \cite{Berezin:1976eg, Casalbuoni:1975hx, Casalbuoni:1975bj}. This was employed to represent a spinning point particle in flat spacetime \cite{Brink:1976uf}, which is described in terms of its position $x^{\mu}(\tau)$ and of an additional spin degree of freedom $\psi_{k}^{\mu}(\tau)$, an odd element of a Grassmann algebra with $\tau$ any parameter along the world line of the particle. 

The $N=2$ model, by allowing the presence of an $SO(2)$ Chern-Simons term, enables us to obtain wave equations of arbitrary $p$-forms in any space-time dimension. Furthermore, the background is necessarily flat for $N>2$ \cite{Howe:1989vn}. It is just in this point that we call attention: in this work we study the confinement of the $N=2$ spinning particle regarding the Randall-Sundrum scenario. We show that the particle is not confined over the membrane, a result that affects the interpretation in terms of $p$-forms and the usual models of field localization, in other words, its non-localization. At the same time, we point out its satellite behavior, a curious effect in extra dimensions in this kind of model, such an effect already shown in \cite{Souza:2019jqz}. 

We organize this work as follows. In the second section, we review the physics of the spinning particle mechanics and its interpretation through antisymmetric tensor fields. In the third section, we discuss the issue of localization giving emphasis to the behavior of some braneworld models. Finally, we present conclusions and perspectives of future  works.

\section{ Particle Mechanics and Antisymmetric Tensor Fields}

In this section we will make a brief review of the spinning particles with spin $N/2$ in flat spacetime, where $N\in\mathbb{Z}_{+}$. First, we will study the motion of these particles, which is described by an action that is invariant under arbitrary reparameterizations of $\tau$, local supersymmetry transformations and local $O(N)$ transformations \cite{Gershun:1979fb}, as we will see below. We briefly review the massless and massive spinning particles, and quantize them, emphasizing the $N=2$ particle model for which a wavefunction with $O(2)$ charge $q$ can be interpreted as
the field strength of a $(q-1)$-form gauge potential by introduction of the Chern-Simons term to break $O(2)$ to $SO(2)$ symmetry. Finally, we will see extensions of this model that yields field equations for massless and massive antisymmetric tensors in arbitrary spacetime dimensions. For more specific details, we call attention to the references cited here.

\subsection{Action for Spinning Particles}

In order to describe spinning particles, we generalize the Lagrangian of the spinless particle
$L=\frac{1}{2}E^{-1}\eta_{\mu\nu}\dot{x}^{\mu}\dot{x}^{\mu}-\frac{1}{2}Em^{2}$,
where $x^{\mu}\equiv x^{\mu}(\tau)$ is the position of the point particle, $E$ is an invariant parameter interpreted as the one-dimensional $\it{vierbein}$ and $m$ is the mass of the particle. It is introduced one additional anticommuting function, the Grassmann variables $\psi^{\mu}_{k}(\tau),\,\,k=1\dots N$, the superpartners of the position $x^{\mu}$, together with a fermionic counterpart $\lambda_{k}$ to $E$, which are parameters invariant and locally supersymmetric \cite{Brink:1976uf, Brink:1976sc, Gershun:1979fb}. Hence, there are enough local symmetries to ensure only physical states in the spectrum. Regarding these aspects, the Lagrangian for a massless $(m=0)$ particle with spin $N/2$ is given by \cite{Gershun:1979fb}.
\begin{eqnarray} 
    L&=&\frac{1}{2}E^{-1}\eta_{\mu\nu}\dot{x}^{\mu}\dot{x}^{\mu}-\frac{i}{2}\eta_{\mu\nu}\psi^{\mu}_{k}\dot{\psi}^{\nu}_{k}-\frac{i}{2}E^{-1}\eta_{\mu\nu}\lambda_{k}\psi^{\mu}_{k}\dot{x}^{\nu}\nonumber\\
    &&-\frac{1}{8}E^{-1}\eta_{\mu\nu}(\lambda_{k}\psi^{\mu}_{k})(\lambda_{k}\psi^{\nu}_{k})+i\eta_{\mu\nu}\psi^{\mu}_{k}\psi^{\nu}_{l}V_{kl},\label{spn1}
\end{eqnarray}
where repeated Latin indices indicate sums. The action related to the Lagrangian above is invariant under reparameterization of the parameter $\tau\rightarrow\tau+s(\tau)$, under supersymmetry and local $O(N)$ transformation \cite{Brink:1976uf, Gershun:1979fb} as shown in the table below 
\renewcommand{\arraystretch}{1.2}
\begin{table}[h]
\centering
\begin{tabular}{|c|l|l|l|}
\hline
\textbf{Variables} & \textbf{Reparametrization} & \textbf{Supersymmetry} & \textbf{Local O(N)} \\
\hline
\hline
$\delta{x^{\mu}}$ & $s\dot{x}^{\mu}$ & $i\alpha_{k}\dot{\psi^{\mu}_{k}}$ & $0$ \\
\hline

$\delta{\psi^{\mu}_{k}}$ & $s\dot{\psi}^{\mu}_{k}$ & $\alpha_{k}\Big[E^{-1}\dot{x}^{\mu}-\frac{i}{2}E^{-1}\lambda_{l}\psi^{\mu}_{l}\Big]$ & $0$ \\ \hline

$\delta{E}$ & $s\dot{E}+\dot{s}E$ & $i\alpha_{k}\lambda_{k}$ & $0$ \\ \hline

    $\delta{\lambda_{k}}$ & $s\dot{\lambda_{k}}+\dot{s}\lambda_{k}$ & $2\dot{\alpha}_{k}+4\alpha_{l}V_{kl}$ & $t_{kl}\lambda_{l}$ \\ \hline

$\delta V_{kl}$ & $s\dot{V}_{kl}+\dot{s}{V}_{kl}$ & $0$ & $\frac{1}{2}\dot{t}_{kl}+t_{kn}{V}_{nl}-t_{ln}{V}_{nk}$ \\ \hline
\end{tabular}
\caption{Symmetry transformations of the spinning particle model \cite{Gershun:1979fb}.}
\label{tab:transform}
\end{table}
\newline
where $\alpha_{k}$ is an odd element of a Grassmann algebra, which is an arbitrary function of $\tau$, and $t_{kl}$ are the generators of the group $O(N)$. The Euler-Lagrange equations obtained by varying the action related to the Lagrangian, equation~(\ref{spn1}), are given by
\begin{eqnarray}
&& p_{\mu}=E^{-1}\left(\eta_{\mu\nu}\dot{x}^{\nu}-\frac{i}{2}\lambda_{k}\eta_{\mu\nu}\psi^{\nu}_{k}\right) \label{motfl1}\\
&&\dot{p}_{\mu}=0\label{motfl2}\\
&&\dot{\psi}^{\mu}_{k}=\frac{\lambda_{k}}{2}p^{\mu}+2\psi^{\mu}_{l}V_{kl},\label{motfl3}
\end{eqnarray}
and the first kind constraints are
\begin{eqnarray}    &&\eta_{\mu\nu}p^{\mu}p^{\nu}\approx0\label{relfl1}\\
&&\eta_{\mu\nu}\psi^{\mu}_{k}p^{\nu}\approx0\label{relfl2}\\
&&\eta_{\mu\nu}\psi^{\mu}_{k}\psi^{\nu}_{l}\approx0\label{relfl3}\\
&&\pi^{\mu}_{k}-\frac{i}{2}\psi^{\mu}_{k}\approx0\label{relfl4},
\end{eqnarray}
where $\pi^{\mu}_{k}=\frac{\partial L}{\partial\dot{\psi}^{\mu}_{k}}$. These equations describe the motion of a free massless spinning particle with spin $N/2$ after quantization.

For a massive spinning particle, it was proposed in \cite{Brink:1976uf} the introduction of an additional (Minkowsky scalar) Grassmann variable $\psi^{5}$ which goes over to $\gamma^{5}$ in the quantization procedure, and this field could carry a mass in the constraint. We further introduced a ``cosmological'' term $\frac{1}{2}E^{-1}m^{2}$ in the Lagrangian to give the mass shell condition. Then, for the action to be invariant, we can extend the Lagrangian, equation~(\ref{spn1}), for a massive case adding the following term to the Lagrangian \cite{Brink:1976uf, Gershun:1979fb} 
\begin{equation}\label{l5flat}
    L_{5}=-\frac{E}{2}m^{2}-\frac{i}{2}\psi^{5}_{k}\dot{\psi}^{5}_{k}-\frac{i}{2}\lambda_{k}m\psi^{5}_{k}+i\psi^{5}_{k}\psi^{5}_{l}V_{kl},
\end{equation}
where
\begin{equation}
    \delta\psi_{k}^{5}=\alpha_{k}{m},
\end{equation}
which is analogous to introducing the ``Goldstone field" with the supersymmetrical transformation law above \cite{Volkov:1973ix}. The complete action for a massive spinning particle, as well as the massless case, equation~(\ref{spn1}), transforms as a total derivative under supersymmetry transformations in table (\ref{tab:transform}).
Then, the set of equations of motion related to the massive spinning particle are equations~(\ref{motfl1}), (\ref{motfl2}) and (\ref{motfl3}).
The first kind constraints are
\begin{eqnarray}    &&\eta_{\mu\nu}p^{\mu}p^{\nu}+E^{2}m^{2}\approx0\label{vinm11}\\
&&\eta_{\mu\nu}\psi^{\mu}_{k}p^{\nu}+m\psi^{5}_{k}\approx0\label{vinm12}\\
&&\eta_{\mu\nu}\psi^{\mu}_{k}\psi^{\nu}_{l}+\psi^{5}_{k}\psi^{5}_{l}\approx0\label{vinm13}\\
&&\pi^{\mu}_{k}-\frac{i}{2}\psi^{\mu}_{k}\approx0\label{vinm14}.
\end{eqnarray}
These equations describe the motion of a massive spinning particle with spin $N/2$ \cite{Gershun:1979fb}.

\subsection{Quantization and \texorpdfstring{$P$}{P}-Forms}

From quantization of point particles models, relativistic free field equations for particles of arbitrary spin can be obtained \cite{Gershun:1979fb, Howe:1989vn, Howe:1988ft}. The basic example is the case with the Klein-Gordon equation, which can be obtained by the quantization of the spinless particle, being the operator $p^{2}$ the generator of reparameterizations of the particle worldline. Similarly, the Dirac equation for massless particles is obtained by quantization of the spinning particle with $N=1$ as in \cite{Berezin:1976eg, Brink:1976uf, Brink:1976sz}
where the Grassmann variable $\psi^{\mu}$ becomes the Dirac matrices on quantization and the generator of supersymmetry becomes the Dirac operator. Then, the action for a massless spinning particle, equation~(\ref{spn1}), moving in a Minkowsky spacetime with coordinates $x^{\mu}$ and momentum $p^{\mu}$ in first-order form is
\begin{equation} \label{actlegm0}   
S=\int \mathrm{d}\tau\left(\eta_{\mu\nu}\dot{x}^{\mu}p^{\nu}+\frac{i}{2}\eta_{\mu\nu}\psi^{\mu}_{k}\dot{\psi}^{\nu}_{k}-\frac{1}{2}E\eta_{\mu\nu}p^{\mu}p^{\nu}-i\lambda_{k}\eta_{\mu\nu}\psi^{\mu}_{k}p^{\mu}-\frac{i}{2}\eta_{\mu\nu}\psi^{\mu}_{k}\psi^{\nu}_{l}f_{kl}\right).
\end{equation}
where, $V_{kl}\equiv\frac{1}{2}f_{kl}$ and the actions above are $O(N)$ invariant. The gauge fields $E$, $\lambda_{k}$ and $f_{kl}$ are Lagrange multipliers that, when implemented as constraints on the particle’s quantum wavefunction, yield a relativistic wave equation for a pure spin $N/2$
particle (in four dimensions) \cite{Howe:1989vn}. The action also has a $d$-dimensional  conformal invariance: these wave equations are the conformal wave equations for arbitrary spin \cite{Siegel:1988ru}. 

Here we emphasize the $N=2$ model, where we can add the action a Chern-Simons term 
\begin{equation}\label{chsi}
    \int dt \epsilon^{ij}f_{ij}
\end{equation} 
which is not possible for $N>2$ once the internal symmetry group is non-abelian. Without the Chern-Simons term, the $N = 2$ wavefunction necessarily must vanish for $D$ odd \cite{Howe:1989vn}. This can be understood in the context of path-integral quantization as a consequence of a global $SO(2)$ anomaly \cite{Howe:1989vn, Rivelles:1990dq}. With the inclusion of the Chern-Simons term with coefficient $(q-\frac{1}{2}D)$ the wavefunction becomes a harmonic $q$-form \cite{Howe:1989vn}.

We shall now consider the $N = 2$ massless case in more detail. The action, equation~(\ref{actlegm0}), can be written as
\begin{eqnarray}
S&=&\int \mathrm{d} \tau\left[\dot{x}^{\mu} p_{\mu}+\mathrm{i}  \eta_{\mu\nu}\bar{\psi}^{\mu} \psi^{\nu}-\mathrm{i} \lambda \bar{\psi}^{\mu} p_{\mu}-\mathrm{i} \bar{\lambda} \psi^{\mu} p_{\mu}-\frac{1}{2} e p^{2}\right.\nonumber\\
&&\left.+f\left(\frac{1}{2}\left[\psi^{\mu}, \bar{\psi}^{\nu}\right] \eta_{\mu \nu}-\left(q-\frac{1}{2} D\right)\right)\right],\label{actlegchm0}
\end{eqnarray}
after we define conveniently the variables
\begin{align}
&\psi^{\mu}=\sqrt{\frac{1}{2}}\left(\psi_{1}^{\mu}+\mathrm{i} \psi_{2}^{\mu}\right),\,\, \bar{\psi}^{\mu}=\sqrt{\frac{1}{2}}\left(\psi_{1}^{\mu}-\mathrm{i} \psi_{2}^{\mu}\right), \\
&\lambda=\sqrt{\frac{1}{2}}\left(\lambda_{1}+\mathrm{i} \lambda_{2}\right),\,\, \bar{\lambda}=\sqrt{\frac{1}{2}}\left(\lambda_{1}-\mathrm{i} \lambda_{2}\right).
\end{align}

This action differs from the specialization given in equation~(\ref{actlegm0}), where $N=2$, through the inclusion of the additional ($q-\frac{1}{2} D$) term \cite{Howe:1989vn}. This addition is consistent with supersymmetry because the supersymmetry variation of $f$ vanishes, as well as the transformation of the $V_{kl}$ (see table~(\ref{tab:transform})). It is also consistent with worldline diffeomorphism, and $\mathrm{SO}(2)$ invariance. This happens because the symmetry $\delta f$ gives a total derivative. For $N>2$ the $\operatorname{SO}(N)$ variation of $f_{ij}$ is not a total derivative, so this modification is no longer possible \cite{Howe:1989vn}.

In the quantization procedure, the $\psi, \bar{\psi}$ variables satisfy the anticommutation relations
\begin{align}
& \left\{\psi^{\mu}, \psi^{\nu}\right\}=\left\{\bar{\psi}^{\mu}, \bar{\psi}^{\nu}\right\}=0 \\
& \left\{\psi^{\mu}, \bar{\psi}^{\nu}\right\}=\eta^{\mu \nu}.\label{antcomut}
\end{align}

Since the $\bar{\psi}^{\mu}$ are a set of mutually anticommuting operators they can be diagonalised on a basis of eigenstates $|\bar{\alpha}\rangle$ for which the eigenvalues $\bar{\alpha}^{\mu}$ of $\bar{\psi}^{\mu}$ are anticommuting. Thus
\begin{equation*}
\bar{\psi}^{\mu}|\bar{\alpha}\rangle=\bar{\alpha}^{\mu}|\bar{\alpha}\rangle .
\end{equation*}
The wavefunction of the $N=2$ particle \cite{Howe:1989vn}, $(\bra{x}\otimes\bra{\alpha})\ket{\Psi}=\Psi(x,\alpha)$, can be expanded as a power series in $\alpha^{\mu}$ in the following way
\begin{align*}
& \Psi=F(x)+\alpha^{\mu} F_{\mu}(x)+\frac{1}{2} \alpha^{\mu} \alpha^{\nu} F_{\mu v}(x)+\ldots \\
& \quad+\frac{1}{p!} \alpha^{\mu_{1}} \ldots \alpha^{\mu_{p}} F_{\mu_{1} \ldots \mu_{p}}(x)+\ldots+\frac{1}{D!} \alpha^{\mu_{1}} \ldots \alpha^{\mu_{D}} F_{\mu_{1} \ldots \mu_{D}}(x). 
\end{align*}
And the constraint imposed by $f$ can be written as
\begin{equation}\label{constf}
    (\eta_{\mu\nu}\psi^{\mu}\bar{\psi}^{\nu}-q)\ket{\Psi}=0,
\end{equation}
by using \eqref{antcomut}. So, the equation \eqref{constf} is equivalent to \cite{Howe:1989vn}
\begin{equation}
    \alpha^{\mu}\frac{\partial}{\partial\alpha^{\mu}}\Psi(x,\alpha)=q\Psi(x,\alpha),
\end{equation}
which is solved by writing $\Psi(x,\alpha)$ as
\begin{equation}
\Psi(x, \alpha)=\frac{1}{q!} \alpha^{\mu_{1}} \ldots \alpha^{\mu_{q}} F_{\mu_{1} \ldots \mu_{q}}(x). 
\end{equation}
The two independent constraints imposed by $\lambda$ and $\bar{\lambda}$ are equivalent to \cite{Howe:1989vn}
\[
{\partial}_{\left[\mu_{1}\right.} F_{\left.\mu_{2} \ldots \mu_{q-1}\right]}=0;\,\ \partial^{\mu_{1}} F_{\mu_{1} \ldots \mu_{q}}=0, 
\]
respectively. The first equation is solved by $F_{\mu_{1} \ldots \mu_{q}}=q \partial_{\left[\mu_{1}\right.} A_{\left.\mu_{2} \ldots \mu_{q}\right]}$ and the second is then the usual field equation for the ($q-1$)th rank antisymmetric tensor gauge potential $A_{\mu_{1}}\ldots \mu_{q-1}$, and it holds for $D\in\mathcal{N}$. 

This mechanism allows us to obtain massive fields for $N=2$ without restrictions  \cite{Howe:1989vn}. Then, for a massive spinning particle, as well as in the massive case, we want to obtain an action with coordinates $x^{\mu}$ and momentum $p^{\mu}$ in first-order form with the Chern-Simons, equation~(\ref{chsi}) term for $N=2$. The action is given by \cite{Howe:1989vn}
\begin{eqnarray}
S&=&\int \mathrm{d} \tau\Bigg[\dot{x}^{\mu} p_{\mu}+\mathrm{i} \eta_{\mu\nu}\bar{\psi}^{\mu} \psi^{\nu}+\mathrm{i}\bar{\psi}^{5} \psi^{5}-\mathrm{i} \lambda (\bar{\psi}^{\mu}p_{\mu}-m\bar{\psi^{5}})-\mathrm{i} \bar{\lambda}(\psi^{\mu} p_{\mu}-m{\psi^{5}})\nonumber\\
&&-\frac{1}{2}e(p^{2}+m^{2})+f\left(\frac{1}{2}\left[\psi^{\mu}, \bar{\psi}^{\nu}\right] \eta_{\mu \nu}+\frac{1}{2}\left[\psi^{5}, \bar{\psi}^{5}\right]-\left(q-\frac{1}{2} D\right)\right)\Bigg],
\label{actlegchm1}
\end{eqnarray}
where $\psi^{5}=\sqrt{\frac{1}{2}}\left(\psi_{1}^{5}+\mathrm{i} \psi_{2}^{5}\right)$. Then, the complex wavefunction satisfying the $SO(2)$ constraint imposed by $f$ can be written as
\begin{equation}
    \frac{1}{q!}\alpha^{\mu_{1}}\dots\alpha^{\mu_{q}}F_{\mu_{1}\dots\mu_{q}}+\frac{im}{(q-1)!}\beta\alpha^{\mu_{1}}\dots\alpha^{\mu_{q}}A_{\mu_{1}\dots\mu_{q-1}}=0
\end{equation}
where $F_{\mu_{1}\dots\mu_{q}}$ and $A_{\mu_{1}\dots\mu_{q-1}}$ are, a priori, also complex, and $\beta$ is the anticommuting eigenvalue of
$\psi^{5}$, such that, $\langle\beta|\psi^{5}=\langle\beta|\beta$. The constraint imposed by $\bar{\lambda}$ yields
\begin{equation}    F_{\mu_{1}\dots\mu_{q}}=q\partial_{[\mu_{1}}{A_{\mu_{2}\dots\mu_{q}]}},
\end{equation}
where $A_{\mu_{1}\dots\mu_{q-1}}$ is the gauge potential for $F_{\mu_{1}\dots\mu_{q}}$. Finally, the constraint related to $\lambda$ gives the antisymmetric tensor generalization of the Proca equation \cite{Howe:1989vn}
\begin{equation}
 \partial^{\mu_{1}}{F_{\mu_{1}\mu_{2}\dots\mu_{q}}} -m^{2}A_{\mu_{1}\mu_{2}\dots\mu_{q-1}}=0. 
\end{equation}

In conclusion, we can observe that the description of the spinning particle for $N=2$ with the Chern-Simons term gives us the description of $p$-forms. In the next section we will analyze the behavior of these particles in a braneworld scenario.

 \section{Confinement of \texorpdfstring{$N=2$}{N} Spinning Particle}

In this section we study the confinement of the spinning particle in the context of Randall-Sundrum scenario in $D=5$ dimensions. Here, the main purpose is to discuss if the superparticle confinement can be interpreted as localization of the related $p$-form fields, since the spinning particle description in curved space-time is possible only for $N\leq2$ \cite{Howe:1988ft}. At the end of the section, we discuss the effective potential for some specific braneworld models.

\subsection{\texorpdfstring{$D=5$}{D=5} Spinning Particle}

In the description of the spinning particle in curved space-time, we can make the minimal coupling $\eta_{AB}\rightarrow{g_{AB}(x)}$ in the action related to the Lagrangian, equation~(\ref{spn1}), with the Chern-Simons term. Then, the action for a $N=2$ spinning particle is given by:
\begin{eqnarray}\label{fullspinac}
    S&=&\int\mathrm{d}\tau\Bigg\{\frac{E^{-1}}{2}g_{AB}\dot{x}^{A}\dot{x}^{B}-\frac{E}{2}m^{2}-\frac{i}{2}g_{AB}\psi^{A}_{k}\frac{D\psi^{B}_{k}}{D\tau}-\frac{i}{2}\psi^{5}_{k}\dot{\psi^{5}_{k}}\Bigg.\nonumber\\
    &-&\Bigg.\frac{i}{2}\lambda^{k}(E^{-1}g_{AB}\psi^{A}_{k}\dot{x}^{B}+m\psi^{5}_{k})-\frac{1}{8}E^{-1}g_{AB}(\lambda_{k}\psi^{A}_{k})(\lambda_{k}\psi^{B}_{k})\Bigg.\nonumber\\    &+&\Bigg.f\left[-i\frac{1}{2}\epsilon_{kl}\Big(g_{AB}\psi^{A}_{k}\psi^{B}_{l}+\psi^{5}_{k}\psi^{5}_{l}\Big)-\Big(q-\frac{1}{2}D\Big)\right]\Bigg\},
\end{eqnarray}
where 
\begin{equation}
    \frac{D\psi^{A}_{k}}{D\tau}=\dot{\psi}^{A}_{k}+\Gamma^{A}_{BC}{\psi}^{B}_{k}\dot{x}^{C}.
\end{equation}
is the covariant derivative of the Grassmann variable $\psi^{A}_{k}$ \cite{Rietdijk:1989qa} and $g_{AB}$ is, initially, a general metric in a $5$-dimensional spacetime. Then, we obtain from equation~(\ref{fullspinac}) the following equations of motion:
\begin{eqnarray}
\frac{\delta S}{\delta E} &\to& g_{AB}\left[\dot{x}^{A}\dot{x}^{B}-i\lambda_{k}\dot{x}^{A}\psi^{B}-\frac{1}{4}(\lambda_{k}\psi^{A}_{k})(\lambda_{k}\psi^{B}_{k})\right]+E^{2}m^{2}=0,\label{402}\\
\frac{\delta S}{\delta \lambda_{k}} &\to & g_{AB}\left[\dot{x}^{A}\psi^{B}-\frac{i}{2}(\lambda_{k}\psi^{A}_{k})\psi^{B}_{k}\right]+Em\psi^{5}_{k}=0,\label{404}\\
\frac{\delta S}{\delta f} &\to& i\frac{1}{2}\epsilon_{kl}(g_{AB}\psi^{A}_{k}\psi^{B}_{l}+\psi^{5}_{k}\psi^{5}_{l})+\Big(q-\frac{1}{2}D\Big)=0\label{409}\\
\frac{\delta S}{\delta \psi^{B}_{k}} &\to& \frac{D\psi^{B}_{k}}{D\tau}-\frac{1}{2}E^{-1}\lambda_{k}\dot{x}^{B}+\frac{i}{4}E^{-1}\lambda_{k}(\lambda_{k}\psi^{B}_{k})+f\epsilon_{lk}\psi^{B}_{l},\label{403}\\
\frac{\delta S}{\delta \psi^{5}_{k}}&\to & \dot{\psi^{5}_{k}}-2\psi^{5}_{l}V_{kl}-\frac{1}{2}\lambda_{k}m=0,\label{405}\\
\frac{\delta S}{\delta x^{A}} &\to& \frac{D}{D\tau}\left[E^{-1}g_{AF}\dot{x}^{F}\right]-\frac{i}{2}\frac{d}{d\tau}\left[\lambda_{k} E^{-1}g_{AF}\psi^{F}_{k}\right]\nonumber\\& &+\frac{i}{2}R_{ASQR}\psi^{Q}_{k}\psi^{R}_{k}\dot{x}^{S}=0.
\label{406}
\end{eqnarray}
We can note that $E, \lambda_k\, \text{and}\, f $ have no dynamics. Therefore, they can be gauged away in the theory. In opposition, the equations (\ref{403})-(\ref{405}) and (\ref{406}) contain the dynamics of the spinning particle. From now on, we will fix the gauges and study the confinement of the spinning particle.

Let us consider the metric $g_{AB}$ in the form of a generic braneworld Randall-Sundrum background with the metric given by
\begin{equation}\label{metric}
ds^{2}=e^{2A(y)}\eta_{\mu\nu}dx^{\mu}dx^{\nu}+dy^{2},
\end{equation}
where, $\eta_{\mu\nu}=\text{diag}(-1,1,1,1)$ and $A(y)$ is the warp factor. If we choose the gauge conditions on equations (\ref{402})-(\ref{409}) as $E=e^{2A(y)}$, $\lambda_{k}=0$ and $f=0$, respectively, we obtain the following equations of motion:
\begin{eqnarray}
&&\frac{D\psi^{P}_{k}}{D\tau}=0,\label{503}\\
&&\dot{\psi^{5}_{k}}=0\\
&&\frac{D}{D\tau}\left[e^{-2A(y)}\dot{x}^{N}\right]+\frac{i}{2}R^{N}_{\ SQR}\psi^{Q}\psi^{R}\dot{x}^{S}=0.\label{506}
\end{eqnarray}
The constraints are given by
\begin{eqnarray*}
&&g_{PQ}\dot{x}^{P}\dot{x}^{Q}+e^{4A(y)}m^{2}=0,\label{502}\\
&&g_{PQ}\dot{x}^{P}\psi^{Q}+me^{2A(y)}\psi^{5}_{k}=0,\label{504}\\
&&g_{AB}\psi^{A}_{k}\psi^{B}_{l}+\psi^{5}_{k}\psi^{5}_{l}=0.
\end{eqnarray*}
By using the metric, equation~(\ref{metric}), we obtain the following Christoffel symbols:
\begin{eqnarray}
    \Gamma^{\rho}_{PQ}&=&[\delta^{\rho}_{P}\delta^{y}_{Q}+\delta^{\rho}_{Q}\delta^{y}_{P}]A'\label{gama1}\\
    \Gamma^{y}_{PQ}&=&-\delta^{\mu}_{P}\delta^{\nu}_{Q}g_{\mu\nu}A'.\label{gama2}
\end{eqnarray}
The Riemann tensor, therefore, is given by:
\begin{eqnarray}
R^{\rho}_{SQR}&=&\delta_{S}^{y}\left[\delta_{R}^{\rho}\delta_{Q}^{y}-\delta_{Q}^{\rho}\delta_{R}^{y}\right]\left[A''+A'^{2}\right]\nonumber\\
&+&\left[\delta_{R}^{\rho}\delta_{Q}^{\mu}-\delta_{Q}^{\rho}\delta_{R}^{\mu}\right]\delta_{S}^{\nu}g_{\mu\nu}A'^{2},\label{412}\\
R^{y}_{SQR}&=&\delta_{S}^{\nu}\left[\delta_{R}^{y}\delta_{Q}^{\mu}-\delta_{Q}^{y}\delta_{R}^{\mu}\right]\left[A''+\!A'^{2}\right]g_{\mu\nu}.\label{413}
\end{eqnarray}
Replacing equations~(\ref{412})-(\ref{413}) in equation~(\ref{506}), and also using the constraint of equation~(\ref{504}) to remove the $x^{\mu}$ dependencies of the equations of motion, we obtain, for $x^{\mu}$ and $y$:
\begin{eqnarray}
\frac{D}{D\tau}\left[e^{-2A}\dot{x}^{\rho}\right]&+&iA''\psi^{y}_{k}\psi^{\rho}_{k}\dot{y}
+im\psi^{\rho}_{k}\psi^{5}_{k} e^{2A}A'^{2}=0,\label{414a}\\
\frac{D}{D\tau}\left[e^{-2A}\dot{y}\right]&-&im\left[A''+A'^{2}\right]e^{2A}\psi^{5}_{k}\psi^{y}_{k}=0.\label{414b}
\end{eqnarray}

Now we solve the equation~(\ref{503}) to find the constants of motion. Using the Christoffel symbols, equations~(\ref{gama1})-(\ref{gama2}), we get
\begin{eqnarray}
\dot{\psi}^{\rho}_{k}+\left[\psi^{y}_{k}\dot{x}^{\rho}+\psi^{\rho}_{k}\dot{y}\right]A'=0,\label{415}\\
\dot{\psi}^{y}_{k}+\psi^{y}_{k}\dot{A}+\psi^{5}_{k} me^{2A}A'=0.\label{419}
\end{eqnarray}
Multiplying the equation~(\ref{419}) by $\psi^{5}_{k}$, we obtain a first-order differential equation with solution
\begin{equation}\label{psitau}
 \psi^{y}_{k}(\tau)\psi^{5}_{k}=\psi^{y}_{(0)k}\psi^{5}_{k}e^{-A}.
\end{equation}
Now, if we multiply both equations by $e^{2A}\psi^{y}_{k}$ and $e^{2A}\psi^{\rho}_{k}$ we can show that
\begin{equation}\label{psi2}
    \frac{d}{d\tau}\left(\psi^{y}_{k}\psi^{\rho}_{k}e^{2A}\right)+\psi^{5}_{k}me^{4A}\psi^{\rho}_{k}A'=0.
\end{equation}
By combining equations~(\ref{gama1}), (\ref{gama2}), (\ref{412}), (\ref{413}), with (\ref{psitau})-(\ref{psi2}) in equations (\ref{414a})-(\ref{414b}) respectively, we get
\begin{eqnarray}
    &&\frac{d}{d\tau}[\dot{x}^{\mu}+i\psi_{k}^{y}\psi_{k}^{\mu}e^{2A}A^{'}]=\frac{dp^{\mu}}{d\tau}=0\label{pmu}\\
    &&\frac{d}{d\tau}[e^{-2A}\dot{y}^{2}+m^{2}e^{2A}-2i\psi_{k}^{5}\psi_{(0)k}^{y}e^{A}A^{'}]\equiv\frac{d}{d\tau}[-\eta_{\mu\nu}p^{\mu}p^{\nu}]=0,\label{p2}
\end{eqnarray}
where we obtain that the conjugate momentum of the spinning particle is conserved in a hypersurface $y=y_{0}$. With the equation~(\ref{p2}) we can construct a conserved equation for the kinetic energy of a spinning particle in extra dimension. Then, comparing $[\eta_{\mu\nu}p^{\mu}p^{\nu}]\Big|_{y=0}=[\eta_{\mu\nu}p^{\mu}p^{\nu}]$, we find
\begin{equation}
    e^{-2A}\dot{y}^{2}=\dot{y}_{0}^{2}-m^{2}\left(e^{2A}-\frac{2i}{m}\psi_{k}^{5}\psi_{(0)k}^{y}e^{A}A^{'}-1\right),
\end{equation}
which gives us directly the effective potential felt by the spinning particle $N\leq2$ for a given warp factor $A(y)$:

\begin{equation}\label{ueff}
   u_{eff}= e^{2A}-\frac{2i}{m}\psi_{k}^{5}\psi_{(0)k}^{y}e^{A}A^{'}-1.
\end{equation}
In the equation above $u_{eff}=\frac{U_{eff}}{m^2}$ is an effective potential density. Now we will analyze the behavior of the $N=2$ spinning particle by looking for its effective potential:
\begin{equation}\label{ueffn2}
   u_{eff}= e^{2A}-\frac{2i}{m}(\bar\psi^{5}\psi^{y}_{(0)}+\psi^{5}\bar\psi^{y}_{(0)})e^{A}A^{'}-1.
\end{equation}

In order to achieve the confinement of the $N=2$ superparticle on the brane, the effective potential must obey the condition $u'_{eff}(0)=0$ and $u''_{eff}(0)>0$. That way the derivative of the effective potential in $y=0$ is given by 
\begin{equation*}
    u'_{eff}(0)=-\frac{2i}{m}(\bar\psi^{5}\psi^{y}_{(0)}+\psi^{5}\bar\psi^{y}_{(0)})A^{''}(0),
\end{equation*}
where, $\psi^{5}=\sqrt{\frac{1}{2}}\left(\psi_{1}^{5}+\mathrm{i} \psi_{2}^{5}\right)$ and $\psi^{y}_{(0)}=\sqrt{\frac{1}{2}}\left(\psi^{y}_{(0)1}+\mathrm{i} \psi^{y}_{(0)2}\right)$. The important point is that the confinement of the test particle can be achieved if $A''(0)=0$. However, for all the braneworld models with localized gravity, the Einstein's equations gives $A''(0)<0$. Therefore, the first condition, necessary to confine $N=2$ spinning particles, is not satisfied. This is enough to already conclude about the non confinement. 

In the next section we will evaluate the behavior of the spinning particle in some usual braneworld models in order to get a better understanding of the effective potential. 

\subsection{Some Braneworld models.}

Now we investigate $u_{eff}$, the effective potential  behavior. We work here by studying the role made by several types of braneworlds found in the literature, basically those with smooth behavior. In other words, they are models supported by scalar fields with kink solutions that mimics the branes in $D=5$.

The most basic models regards the Sine-Gordon and $\phi^{4}$ potentials in the context of Randall-Sundrum background to represent the correspondent smooth brane scenarios. For the Sine-Gordon membrane case, see \cite{Gremm:1999pj} for instance, the warp factor is given by
\begin{equation}
    A(y)=\ln\left[\sinh^{b}{(cy)}\right].
\end{equation}

For the $\phi^{4}$ membrane case \cite{Kehagias:2000au} the warp factor is given by
\begin{equation}
    A(y)=-b\ln\left[\cosh^{2}{(cy)}\right]-\frac{b}{2}\tanh^{2}{(cy)}.
\end{equation}
In the last two cases, $b$ and $c$ are parameters that change the membrane's profile.
The Bloch brane model \cite{Bazeia:2004dh} works with the presence of two real scalar fields as thin brane generators and has $r$ as a parameter to change the profile of solutions. In this case the warp factor is given by
\begin{equation}
    A(y)= \frac{1}{9r} \left[ (1 - 3r) \tanh^2(2ry) - 2 \ln \cosh(2ry) \right].
\end{equation}

The deformed brane solution \cite{Bazeia:2003aw,Cruz:2009kb} implements a deformation in usual scalar field models producing new solutions. In this case, the deformation parameter $s$ is important in the new solutions. For this last one, the warp function is given as

\begin{align}
A_s(y) &= -\frac{1}{3}\frac{s}{2s+1}\tanh^{2s}\left(\frac{y}{s}\right) \nonumber\\
&\quad -\frac{2}{3}\left(\frac{s^2}{2s-1}-\frac{s^2}{2s+1}\right)\left\{\ln\left[\cosh\left(\frac{y}{s}\right)\right]-\sum_{n=1}^{s-1}\frac{1}{2n}\tanh^{2n}\left(\frac{y}{s}\right)\right\}.
\end{align}

\begin{figure}[!ht]
    \centering
    \includegraphics[width=0.49
    \textwidth]{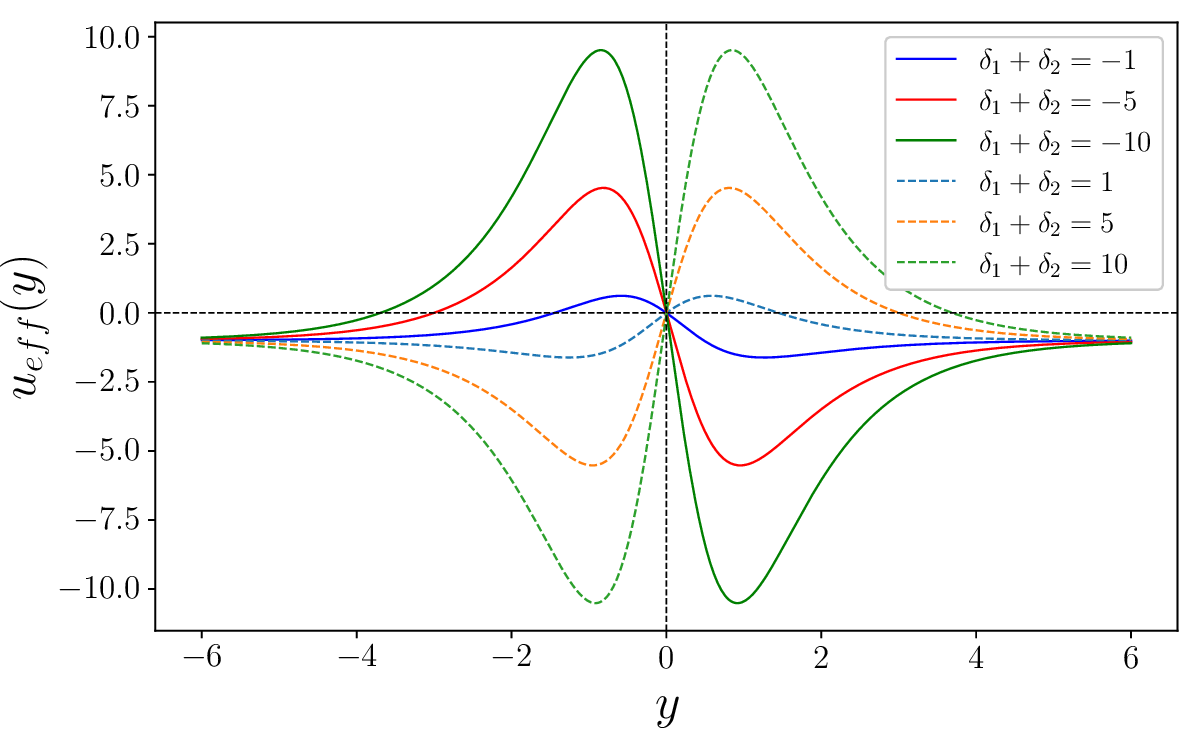}
    \includegraphics[width=0.49
    \textwidth]{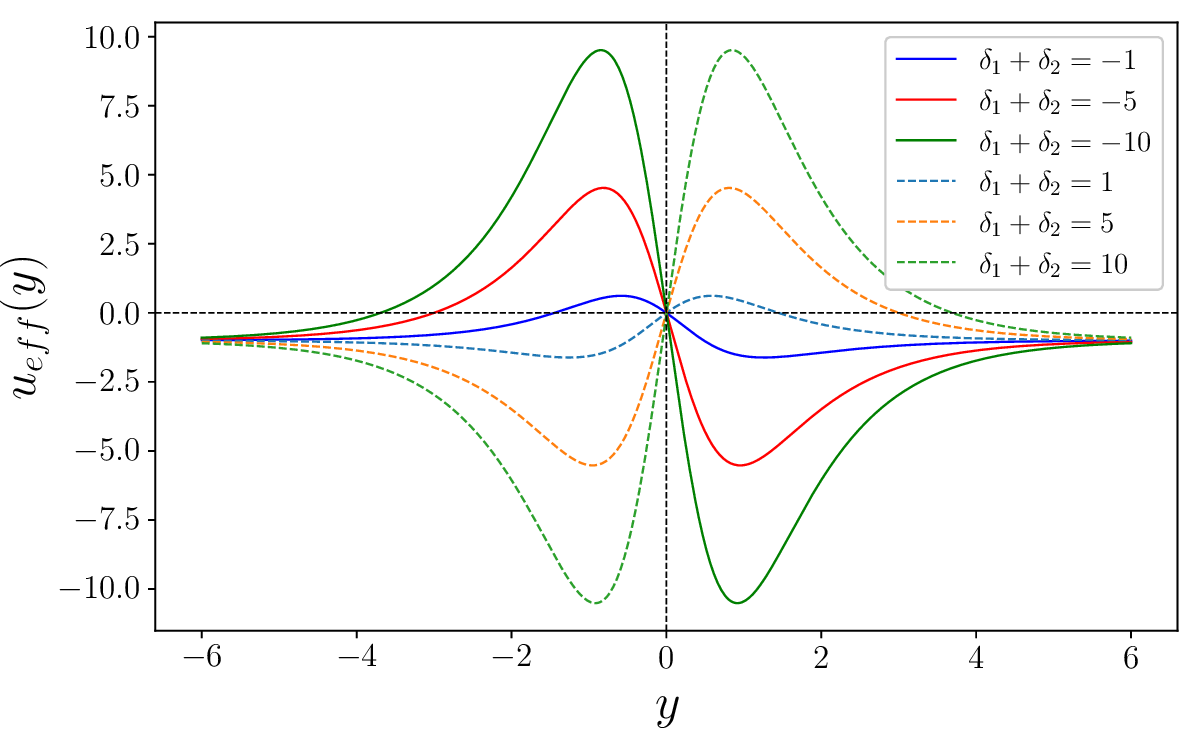}
    \caption{The effective potentials due to the Sine-Gordon (left side) and $\phi^{4}$ (right side)  membrane models.}
    \label{fig:sinegordon}
\end{figure}

\begin{figure}[!ht]
    \centering
    \includegraphics[width=0.9
    \textwidth]{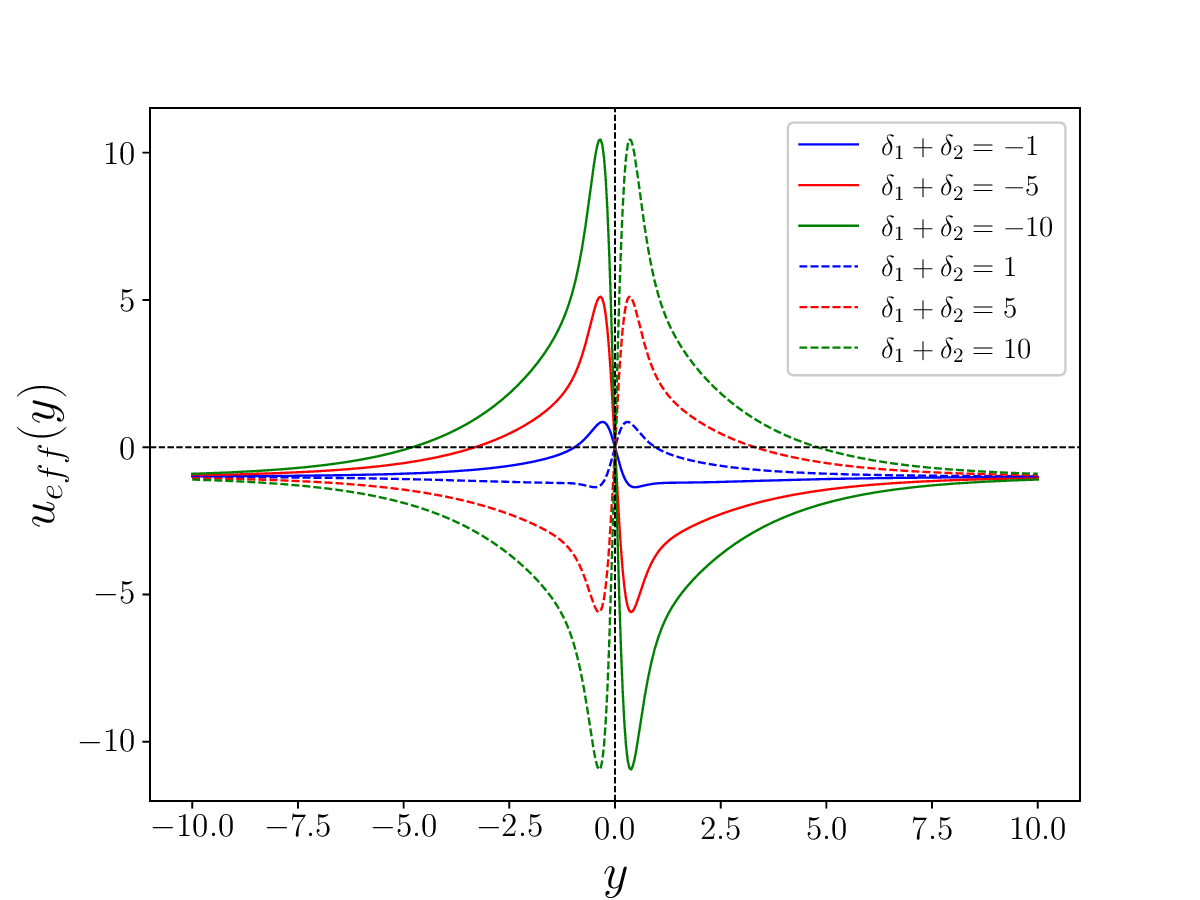}
    \caption{The effective potential due to the Bloch membrane model.}
    \label{fig:bloch}
\end{figure}

\begin{figure}[!ht]
    \centering
    \includegraphics[width=0.49
    \textwidth]{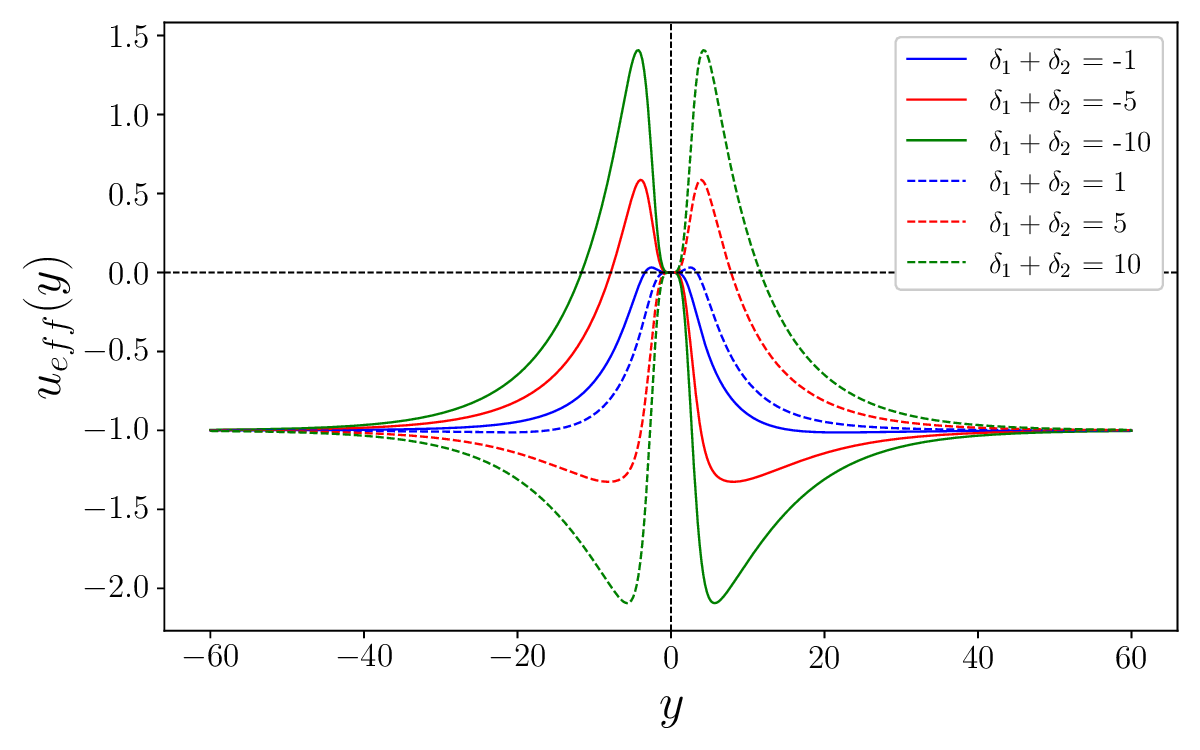}
    \includegraphics[width=0.49
    \textwidth]{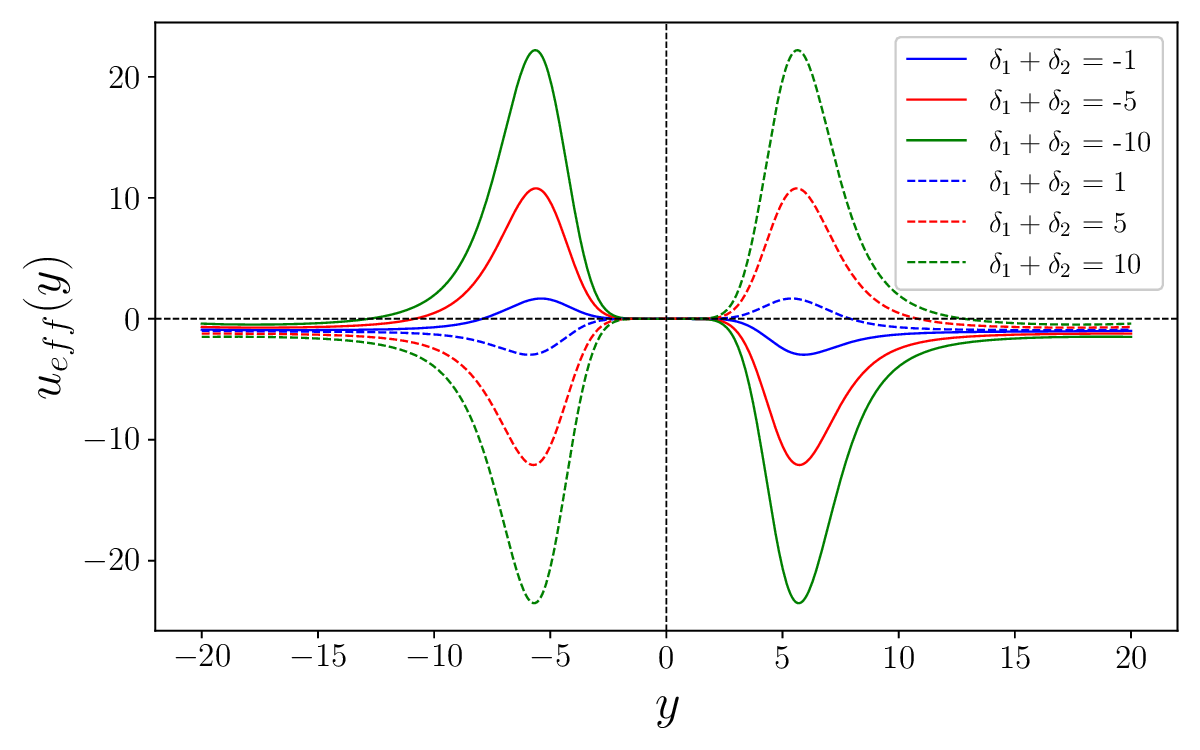}
    \caption{The effective potentials due to the Deformed membrane model: on the left side we plot the potential for $s=3$ and,  on the right side, for $s=5$. In this last case, we start to see the appearance of a flat potential region around $y=0$.}
    \label{fig:def35}
\end{figure}

\begin{figure}[!ht]
    \centering
    \includegraphics[width=0.9
    \textwidth]{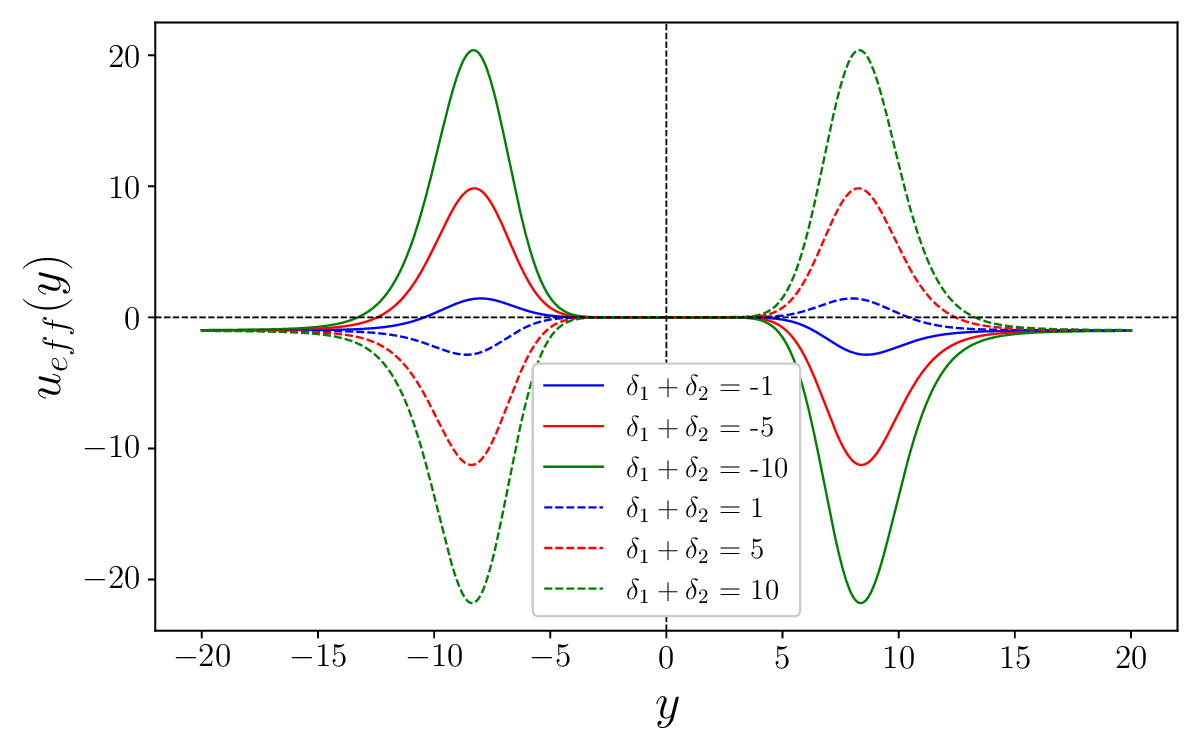}
     \caption{The effective potential due to the Deformed membrane case with $s=7$. Here clearly the flat region shows up. In this case, the satellite particle goes to the left far from $y=0$.}
    \label{fig:def7}
    \end{figure}

For all these cases the profile of the effective potential is quite the same - see figs. (\ref{fig:sinegordon}), (\ref{fig:bloch}), (\ref{fig:def35}), and (\ref{fig:def7}). The constant $\delta_1+\delta_2$, that assuming positive and negative values, plays the role of $\bar\psi^{5}\psi^{y}_{(0)}+\psi^{5}\bar\psi^{y}_{(0)}$ inside the effective potential. We normalize to the parameters $b$, $c$ and $r$ all to $1$, but effectively their influence is to narrow or widen the plot profiles. We can see a potential well with minimum to the left/right of the origin $y=0$, the place where the membrane sits. And there is a potential barrier to the left/right of $y=0$. The model that has some different aspects is the one related to the deformed brane case. Basically, the deformation procedure also goes to the effective potential and, for instance, for $s=5$ and $s=7$, we can see a flat potential region around $y=0$. This is interesting because there is no force over the particle just where we should see the membrane. However, this is not correct: if we study the solutions in \cite{Bazeia:2003aw}, we see that the deformation gives a way to simulate a two membrane solution. When we increase the $s$ values we also increase the distance between these two solutions, as can be seen in figs. $2$ and $4$ of \cite{Bazeia:2003aw}. This explain the flat region presented in fig. (\ref{fig:def7}).

We can analyze the confinement of the $N=2$ superparticle graphically  where the effective potential allows us, for an energy less than the height of the potential well, to localize the spinning particles in a position close to the membrane, but not over the membrane. In fact, the $N=2$ superparticle is not confined over the membrane and, therefore, the $p$-form interpretation is affected: we can not, through this model, localize these fields on the membrane. Despite this, we point out the curious and true satellite behavior of the superparticle, one interesting effect on physics of extra dimensions. This resembles gravitational models to describe planetary orbits around stars and particle orbits around black holes, for instance \cite {Weinberg:1972kfs}.

\section{Conclusions and Perspectives}

In this manuscript, we analyze the behavior of the  $N=2$ spinning particle in a braneworld scenario. This particle, in principle, describe $p$-forms in this background, since in Minkowsky spacetime the wavefunctions obtained by quantization describe $p$-forms in $N=2$ extended supersymmetry \cite{Howe:1989vn}. In principle, we can recover the flat model by taking the action, equation (\ref{fullspinac}), in the locus ($y=0$) of the membrane. Basically we take the action for a spinning particle with a Chern-Simons term and we construct the effective potential in extra-dimensions to study its confinement. We show that, even for the $N=2$ case,  for all the smooth  models with localized gravity, the Einstein's equations gives $A''(0)<0$. Therefore, the first condition necessary to confine $N=2$ spinning particles is not satisfied. This is enough to already conclude about the non confinement. 

We analyzed the effective potential graphically for the Sine-Gordon, $\phi^4$, Bloch and deformed membrane models. The generated plots given have the profile where the effective potential allows us to localize the spinning particles in a position close to the brane, but not over the brane, then establishing a characteristic of non confinement. Furthermore, this result affects directly the $p$-form interpretation and non localization of the fields on the membrane. This result calls for an answer: which model should we in fact consider to treat $p$-form fields? Here we see that it should not be that of $N=2$ superparticle. Despite these facts, we point out the curious satellite like behavior presented by the particle,
since when confined in this potential well, it must orbit the membrane.

It is interesting to study with more care the details of particle scattering with the effective potential found in this paper. In this case, resonances and tunneling effects can be discussed. A nice related problem would be the investigation of the $N=4$ superparticle. In this case, the quantization give us an interpretation of linearized gravity which can be put in the context of Randall-Sundrum scenario again, if we want to study gravity localization in the superparticle viewpoint. These are themes for future works.

\acknowledgments

    I. M. Macêdo and F. E. A. de Souza are thankful for the financial support provided by the Fundação Cearense de Apoio ao Desenvolvimento Científico e Tecnológico (FUNCAP) through processes n{$^\circ$} 31052.000190/2025-40 and FPD-$0213$-$00349.01.01/23$, respectively. The authors are grateful to Prof. Gilberto Dantas Saraiva for useful conversations.


\bibliographystyle{JHEP}
\bibliography{biblio}

\end{document}